\documentclass[prl,aps,floats,twocolumn,floatfix,showpacs]{revtex4}
\usepackage{epsfig}

\input{tcilatex}
\begin{document}
\author{David F. Mota}
\thanks{ DFM thanks Oxford University for hospitality, C.
van de Bruck and M. Murphy for discussions and Funda\c{c}\~{a}o para a Ci%
\^{e}ncia e a Tecnologia, and Funda\c{c}\~{a}o Calouste Gulb\^{e}nkian for
funding.}
\author{John D. Barrow}
\affiliation{DAMTP, Centre for Mathematical Sciences, Cambridge University, Wilberforce
Road, Cambridge CB3 0WA, UK}
\title{Varying Alpha in a More Realistic Universe}
\begin{abstract}
We study the space-time evolution of the fine structure constant, $\alpha$, inside
evolving spherical overdensities in a lambda-CDM Friedmann
universe using the spherical infall model. We show that its value inside
virialised regions will be significantly larger than in the low-density
background universe. The consideration of the inhomogeneous evolution of the
universe is therefore essential for a correct comparison of extragalactic
and solar system limits on, and observations of, possible time
variation in $\alpha$ and other constants. Time variation in $\alpha$ 
in the cosmological background can give rise to no
locally observable variations inside virialised overdensities like the one
in which we live, explaining the discrepancy between astrophysical and
geochemical observations. 
\end{abstract}
\maketitle
Recent observations of small variations in relativistic atomic structure in
quasar absorption spectra \cite{murphy} suggest that the fine structure
constant $\alpha $, was smaller at redshifts $z=1-3$ than the current
terrestrial value $\alpha _{0}=7.29735308\times 10^{-3}$, with $%
\Delta \alpha /\alpha \equiv \{\alpha (z)-\alpha _{0}\}/\alpha
_{0}=-0.543\pm 0.116\times 10^{-5}$. Several theories of varying $\alpha $
have been proposed to investigate the implications \cite{bsbm}. Time-varying 
$\alpha $ requires a scalar field that couples to
electromagnetically-charged matter. Variations in $\alpha $, allowed by the
conservation of energy and momentum will then depend on the expansion of the
universe and the evolution of the electromagnetically coupled matter. The
inclusion of electroweak or grand unification will create more complicated
consequences and constraints \cite{lang}. 
Variations in $\alpha $ due to perturbation in the matter fields were first
studied in \cite{mota2}, using the linear theory of cosmological
perturbations. It was shown there, that perturbations in $\alpha $ will grow
during the matter-dominated epoch and will be constant or decay during the
other eras. In \cite{otoole} the effects of $\alpha $ on gravity were also
analysed. However, any effects on the time and spatial evolution of $\alpha $
due to cluster and galaxy formation has been ignored in the
literature. 
This
is a major weakness and, as a result, past attempts to confront observations
on extragalactic scales with lab or solar system constraints on $\alpha $
variation are all questionable. This letter describes the first attempts to
follow the inhomogeneous evolution of $\alpha $ during the development of
non-linear cosmic structures. We compare the evolution of $\alpha $ in the
background universe and in overdense regions of the universe.

We will study variations in $\alpha $ in the
Bekenstein-Sandvik-Barrow-Magueijo (BSBM) theory \cite{bsm1}, which assumes
that the total action of the Universe is given by: 
\begin{equation}
S=\int d^{4}x\sqrt{-g}\left( \mathcal{L}_{g}+\mathcal{L}_{m}+%
\mathcal{L}_{\psi }+\mathcal{L}_{em}e^{-2\psi }\right)  \label{S}
\end{equation}%
In the BSBM varying-$\alpha $ theory, the quantities $c$ and $\hbar $ are
constants, while $e$ varies as a function of a real scalar field $\psi ,$
with $e=e_{0}e^{\psi }$. $\mathcal{L}_{\psi }={\frac{\omega }{2}}\partial
_{\mu }\psi \partial ^{\mu }\psi $, $\omega $ is a coupling constant, and $%
\mathcal{L}_{em}=-\frac{1}{4}f_{\mu \nu }f^{\mu \nu }$. The gravitational
Lagrangian is the usual $\mathcal{L}_{g}=-\frac{1}{16\pi G}R$, with $R$ the
Ricci curvature scalar. $\mathcal{L}_{m}$ is the matter fields
Lagrangian. 
Defining an auxiliary gauge potential by $%
a_{\mu }=\epsilon A_{\mu }$ and a new Maxwell field tensor by $f_{\mu \nu
}=\epsilon F_{\mu \nu }=\partial _{\mu }a_{\nu }-\partial _{\nu }a_{\mu }$,
the covariant derivative takes the usual form, $D_{\mu }=\partial _{\mu
}+ie_{0}a_{\mu }$. The fine structure constant is then $\alpha \equiv
\alpha _{0}e^{2\psi }$ with $\alpha _{0}$ the present value here.

The background universe will be described by a flat, homogeneous and
isotropic Friedmann metric with expansion scale factor $a(t)$. Varying the
total Lagrangian we obtain the Friedmann equation ($\hbar =c\equiv 1$) for a
universe containing dust, of density $\rho _{m}$ $\propto a^{-3}$ and a
cosmological constant $\Lambda $ with energy-density $\rho _{\Lambda }\equiv 
$ $\Lambda /(8\pi G)$: 
\begin{equation}
3H^{2}=8\pi G\left( \rho _{m}\left( 1+\left\vert \zeta \right\vert
e^{-2\psi }\right) \ +\rho _{\psi }+\rho _{\Lambda }\right)  \label{fried}
\end{equation}%
where $H\equiv \dot{a}/a$ is the Hubble rate, $\rho _{\psi }=\frac{\omega }{2%
}\dot{\psi}^{2},$ and $\zeta =\mathcal{L}_{em}/\rho _{m}$ is the fraction of
the matter which carries electric or magnetic charges. The value (and sign)
of $\zeta $ will depend on the nature of dark matter: $\zeta \approx 10^{-4}$
for neutrons or protons but $\zeta =-1$ for superconducting cosmic strings 
\cite{bsm1}. The scalar-field evolution equation is 
\begin{equation}
\ddot{\psi}+3H\dot{\psi}=-2 e^{{-2\psi }}\zeta \rho _{m}/\omega.
\label{psidot}
\end{equation}

Variations in $\alpha$, due to the formation of overdense spherical
regions, can be studied using the spherical infall model
\cite{padmanaban}. We model the spherical overdense inhomogeneity by a
closed Friedmann metric and the background universe by a spatially flat
Friedmann model. Each has their own expansion scale factor and time, which
are linked by the condition of hydrostatic support at the boundary. This
approach is equivalent to study the effect of perturbations to the
Friedmann metric by considering spherically symmetric regions of different
spatial curvature in accord with Birkhoff's theorem.  The density
perturbations need not to be uniform within the sphere: any spherically
symmetric perturbation will evolve within a given radius in the same way
as a uniform sphere containing the same amount of mass. Clearly, this
model ignores any anisotropic effects of gravitational instability or
collapse. Similar results could be obtained by performing the analysis of
the BSBM theory using a spherically symmetric Tolman-Bondi metric for the
background universe with account taken for the existence of the pressure
contributed by $\Lambda$ and the $\psi $ field. In what follows,
therefore, density refers to \textit{mean} density inside a given sphere.

Consider a spherical perturbation with constant internal density which, at
an initial time, has an amplitude $\delta _{i}>0$ and $|\delta _{i}|\ll 1$. 
At early times the sphere expands along with the background. For a
sufficiently large $\delta _{i},$ gravity prevents the sphere from
expanding. Three characteristic phases can then be identified. \textit{
Turnaround}: the sphere breaks away from the general expansion and reaches a
maximum radius. \textit{Collapse}: if only gravity is significant, the
sphere will then collapse towards a central singularity where the densities
of the matter fields would formally go to infinity. In practice, pressure
and dissipative physics intervene well before this singularity is reached
and convert the kinetic energy of collapse into random motions. \textit{
Virialisation}: dynamical equilibrium is reached and the system becomes
stationary: there are no more time variations in the radius of the system, $%
R $, or in its energy components.
In  the spherical collapse model, due to its symmetry, the only independent coordinates are
the radius of the overdensity and time.  Also, as is standard practice
when using this model, we consider that there are no shell-crossing 
which implies mass conservation inside the overdensity and independence of
the radius coordinate \cite{padmanaban}. The equations can then be written ignoring the
spatial dependence of the fields (but still including an equation for the
evolution of the radius). Hence,
the evolution of a spherical overdense
patch of scale radius $R(t)$ is given by the Friedmann acceleration
equation: 
\begin{equation}
3\ddot{R}=-4\pi G R\left( \rho _{cdm}\left( 1+\left\vert
\zeta \right\vert e^{-2\psi _{c}}\right) +4\rho _{\psi _{c}}-2\rho _{\Lambda
}\right)  \label{rcluster}
\end{equation}%
where $\rho _{cdm}$ is the total density of cold dark matter and baryons in
the cluster, $\psi_c$ is the homogeneous field inside the cluster and we
have used the equations of state $p_{\psi _{c}}=\rho _{\psi _{c}}$, $%
p_{cdm}=0$ and $p_{\Lambda }=-\rho _{\Lambda }$. 
In the cluster, the
evolution of $\psi _{c}$ and $\rho _{cdm}$ is given by 
\begin{equation}
\ddot{\psi _{c}}+3\left(\dot{R}/R\right)\dot{\psi _{c}}=-2
e^{-2\psi _{c}}\zeta \rho _{cdm}/\omega  \label{psidotcluster}
\end{equation}%
and $\rho _{cdm}\propto R^{-3}.$ The cluster will virialise when 
\begin{equation}
T_{vir}=-0.5U_{G}+U_{\Lambda }-2U_{\psi _{c}}  \label{virial}
\end{equation}
where $U_{G}$ $=-\frac{3}{5}GM^{2}R^{-1}$ is the potential energy associated
with the uniform spherical overdensity, $U_{\Lambda }$ $=-\frac{4}{5}\pi
G\rho _{\Lambda }MR^{2}$ is the potential associated with $\Lambda ,$ and $%
U_{\psi _{c}}=-\frac{3}{5}GMM_{\psi _{c}}R^{-4}$ is the potential associated
with $\psi _{c}.$ $T_{vir}$
is the kinetic energy, and $M=M_{cdm}+M_{\psi _{c}}$ is the cluster mass,
with $M_{cdm}=\frac{4\pi }{3}\rho _{cdm}(1+|\zeta |e^{-2\psi _{c}})R^{3}$
and $M_{\psi _{c}}=\frac{4\pi }{3}\rho _{\psi _{c}}R^{6}$; 
Using the virial theorem (\ref{virial}) and energy conservation when the
cluster virialises gives 
\begin{eqnarray}
0.5U_{G}+2U_{\Lambda }-U_{\psi _{c}}\,\,\vert_{z=z_v}
=U_{G}+U_{\Lambda } +U_{\psi _{c}}\,\,\vert_{z=z_{ta}}
\label{virialcond}
\end{eqnarray}
where $z_{v}$ is the redshift of virialisation and $z_{ta}$ is the redshift
at the turn-around of the over-density at its maximum radius, when $%
R=R_{max} $ and $\dot{R}\equiv 0$.

The analysis in this letter (which reduces completely to the linear
perturbation analysis in the small time limit when all inhomogeneities are
small \cite{mota2}) is valid for spherically symmetric perturbations right into the
non-linear regime, including turnaround and collapse. Spatial gradients in
pressure (and hence in the scalar field) are negligible on large scales,
exceeding the Jeans length. The scalar field is indeed slightly
inhomogeneous  but the inclusion of a varying $\alpha $ at a level consistent with observation 
\cite{murphy}, ($\dot{\alpha}/\alpha _{0}\sim 10^{-6}H$), does not affect
the overall expansion of the universe or the dynamical collapse of the
overdense regions. The energy density associated with $\psi $ is always a
negligible contribution to the energy content of the universe, both in the
cluster and in the background, if we are far from the initial or collapse
singularities \cite{bsm1}.
After the fluctuation virialises and attains
gravitational equilibrium this assumption may no longer be true, but our
analysis like that of all other exact studies of galaxy formation breaks
down at that stage because the resulting disk will be controlled by
pressure and gradients. 

\begin{figure}[tbph]
\epsfig{file=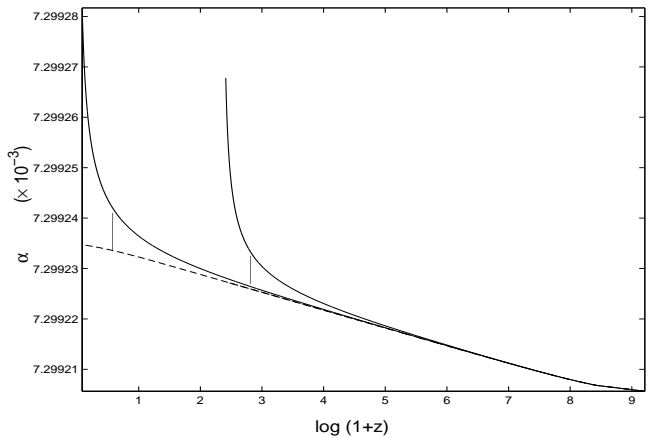,height=6cm,width=8.6cm} %
\caption{{\protect\small {\textit{Evolution of $\protect\alpha $ in the background (dashed line) and inside
clusters (solid lines) as a function of $\log (1+z)$. Initial conditions were
set to match observations of }$\protect\alpha $ variation in ref. 
\protect\cite{murphy}. Two clusters virialise at different
redshifts, one of them in order to have $%
\protect\alpha _{c}(z_{v}=1)=\protect\alpha _{0}$. Vertical lines represent the moment of turn-around.}}}
\label{alphabalphac}
\end{figure}

\begin{figure}[tbph]
\epsfig{file=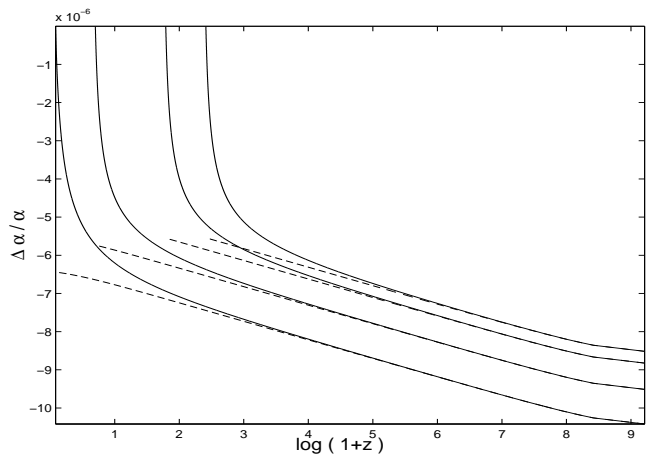,height=6cm,width=8.6cm}
\caption{{\protect\small {\textit{Evolution of $\Delta 
\protect\alpha /\protect\alpha $ in the background (dashed lines) and inside
clusters (solid lines) as a function of $\log (1+z)$. Initial conditions were
set to match observations of }$\protect\alpha $\ variation in ref. 
\protect\cite{murphy}. Four clusters
that virialised at different redshifts. All clusters were started so as to
have $\protect\alpha _{c}(z_{v})=\protect\alpha _{0}$.}}}
\label{deltaalphabalphac}
\end{figure}

The behaviour of the fine structure constant
during the evolution of a cluster can now be obtained by numerically
evolving the background eqs.(\ref{fried})-(\ref{psidot}) and the cluster
eqs.(\ref{rcluster})-(\ref{psidotcluster}) until the virialisation condition
(\ref{virialcond}) holds. 
Since the Earth is in a virialised overdense region, the initial conditions
for $\psi $ are chosen so as to obtain our measured lab value of $\alpha $
at virialisation $\alpha _{c}(z_{v})=\alpha _{0}$ and to match the latest
observations \cite{murphy}
for overdense regions at $3\geq |z-z_{v}|\geq 1$.
Since $z_{v}$ is uncertain,
we have chosen representative examples with virialisation over the range $%
0<z_{v}<10$. This is shown in Fig. \ref{deltaalphabalphac}, where the clusters
have different initial conditions in order to satisfy the constraints
described. This is just an example, since in reality, the initial condition
for $\psi $ needs to be fixed only once, for our Galaxy. Hence, $\alpha $ in
other clusters will have a lower or higher value (with respect to $\alpha
_{0}$) depending on their $z_{v}$; see Fig.\ref{alphabalphac}. After
virialisation occurs the cluster radius becomes constant; time and space
variations in $\alpha $ are suppressed, and $\alpha $ becomes constant. If
there were any variations in $\alpha $ after virialisation, the energy and
radius of the cluster would need to vary in order to conserve energy and
this is inconsistent with virialisation. This phenomenon is not included in
Fig. \ref{alphabalphac}, since we did follow the evolution to virialisation
with a many-body simulation which would need to include the fluid equations
that describe the pressure inside the cluster. In our simulation, the virial
condition is a 'stop condition' and so we just observe the typical behaviour
of the cluster's collapse as $R\rightarrow 0$. Clearly, collapse will never
occur in practice; dissipative physics will eventually intervene and convert
the kinetic energy of collapse into random motions. In addition to the
stationarity condition that occurs when the cluster virialises, it can be
seen from Fig.\ref{deltaalphabalphac} that in all cases the variation in $\alpha $
since the beginning of the cluster formation is of order $10^{-5}$, and
numerical simulations give $\dot{\alpha}/\alpha \approx 10^{-22}s^{-1}.$ If
variations in $\alpha $ are so small for such a wide range of virialisation
redshifts, we can assume that the difference between the value of $\alpha
_{c}$ at $z_{v}$ and at $z=0$ will be negligible. Therefore it is a good
approximation to assume that the time-evolution of both $\alpha _{c}$ and of
the cluster will cease after virialisation. Although this is not necessarily
true (the cluster could keep accreting mass), it is a good
approximation in respect of the evolution of $\alpha $, especially for
clusters which have virialised at lower redshifts. 
%

\begin{figure}[htbp!]
\epsfig{file=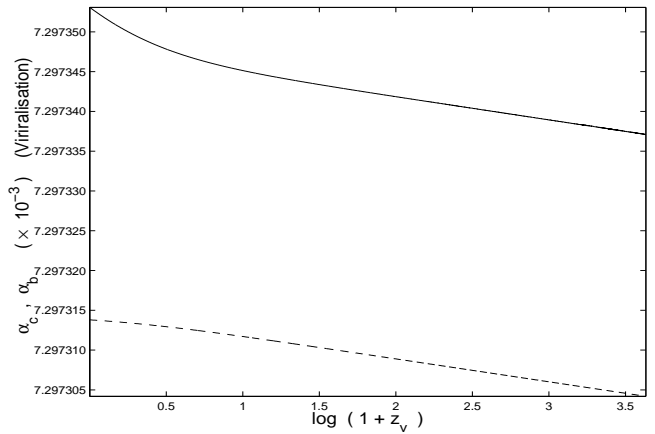,height=6cm,width=8.6cm}
\caption{{\protect\small {\textit{Plot of $\protect\alpha $ 
as a function of $\log (1+z_v)$, at virialisation. Clusters
(solid line), background (dashed line).}}}}
\label{alphabalphav}
\end{figure}

\begin{figure}[hbtp!]
\epsfig{file=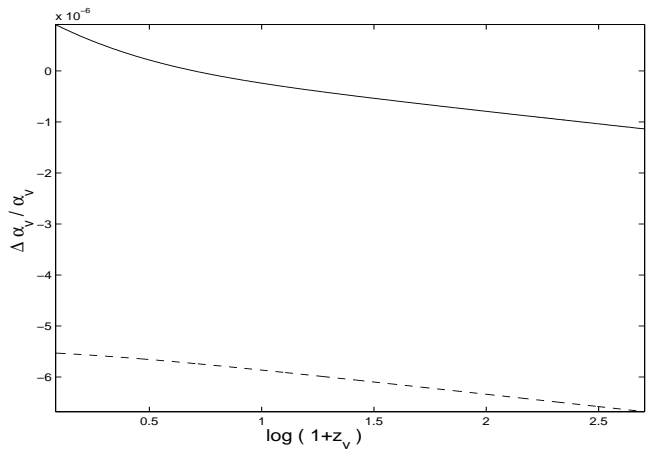,height=6cm,width=8.6cm}
\caption{{\protect\small {\textit{Plot of $\Delta 
\protect\alpha/ \protect\alpha $ as a function of $\log (1+z_v)$, at
virialisation. Clusters
(solid line), background (dashed line). }}}}
\label{deltaalphabalphav}
\end{figure}

If we could measure $\alpha$ inside virialised overdensities and
the corresponding value in the background, at the time of
their virialisation, then Fig \ref{alphabalphav} and
\ref{deltaalphabalphav} would give us the evolution of $\alpha_v$ or
its time shift, $\Delta \alpha_v /\alpha_v ,$ as a function of
$z_v$. Those figures show us that differences in
between the background and the overdensities increase as $z_v\rightarrow0$.
%
%
This is due to the
earlier freezing of the value of $\alpha $ at virialisation, and to our
assumption that we live in a $\Lambda CDM$ universe. \textit{At lower
redshifts, specially after }$\Lambda $\textit{\ starts to dominate,
variations in }$\alpha $\textit{\ in the background are turned off by the
accelerated expansion. However, the value of }$\alpha $\textit{\ in the
collapsing cluster will keep growing until virialisation occurs}. Numerical simulations
give $\Delta\alpha/\alpha (z=2)=-6\times10^{-6}$ for the background 
and $\Delta\alpha/\alpha (z_v=0.13)=-1.5\times10^{-7}$ in a cluster.   
At higher redshifts, $z\gg 1$, both $\alpha _{b}$ and $
\alpha _{c}$ evolve in expanding environments: their increase is logarithmic
in time before $\Lambda $ starts to dominate, so the difference between them
will be much smaller. 
%

\begin{figure}[htbp!]
\epsfig{file=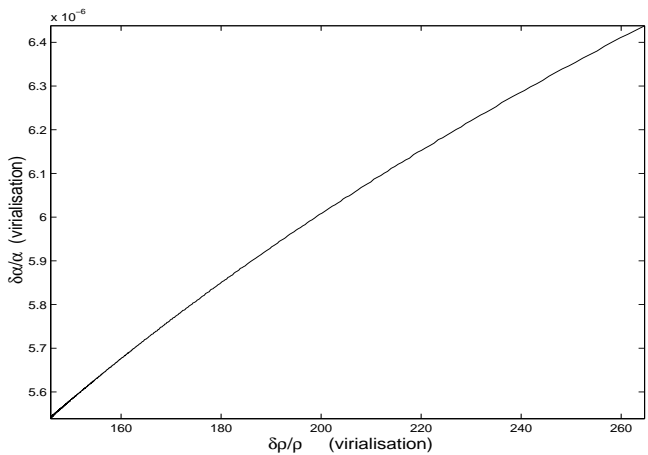,height=6cm,width=8.6cm}
\caption{{\protect\small {\textit{Variation of $\protect\delta \protect%
\alpha /\protect\alpha $ with $\protect\delta \protect\rho /\protect\rho $
at the cluster virialisation redshift. The evolution of $\protect\alpha $
inside the clusters was normalised to satisfy the latest time-variation
observational results, and to have $\protect\alpha _{c}(z_{v})=\protect%
\alpha _{0}$ for }${z}_{v}{=1.}$}}}
\label{alpharho}
\end{figure}

Spatial variations in $\alpha $ can be tracked using a 'spatial' density
contrast, defined by $\delta \alpha /\alpha \equiv \lbrack \alpha
_{c}-\alpha _{b}]/\alpha _{b}$. 
A plot of the spatial inhomogeneities in $\alpha $ with respect to the
matter density inhomogeneity, $\delta \rho /\rho ,$ at virialisation is
shown in Fig. \ref{alpharho}. Note that $\delta \alpha /\alpha $ grows in
proportional to the density contrast of the matter inhomogeneities ($\propto
t^{2/3}$ when both are small during dust domination \cite{mota2}). In a $%
\Lambda CDM$ model, the density contrast, $\Delta _{c}=\rho
_{cdm}(z_{v})/\rho _{b}(z_{v})$ increases as the redshift decreases. For
high redshifts, the density contrast at virialisation becomes the
asymptotically constant in standard ($\Lambda =0$) CDM, $\Delta _{c}\approx
178$ at collapse or $\Delta _{c}\approx 148$ at virialisation. This is
another reason why at lower redshifts the difference between
$\alpha_v$ and $\alpha_b$ increases. Trends
of variation in $\alpha $ can be then predicted from the value of the matter
density contrast of the regions observed. Useful fitting formulas for the
dependence of $\alpha $ variation on $\delta \rho /\rho $ and the scale
factor $a$ $\equiv (1+z)^{-1}$ are:%
\begin{eqnarray}
\delta \alpha /\alpha  &=&(5.56-0.7\sigma ^{\frac{1}{2}}+0.078\sigma
+0.00352\sigma ^{\frac{3}{2}})\times 10^{-6}  \nonumber  \label{fit2} \\
\delta \alpha /\alpha  &=&(5.37+0.373\theta ^{\frac{1}{6}%
}-0.27\theta ^{\frac{1}{3}}+0.007\theta ^{\frac{1}{2}})\times 10^{-6} 
\nonumber  \label{fit3} \\
\Delta \alpha/\alpha  &=&-(2.47-1.81\sigma ^{\frac{1}{2}%
}+0.59\sigma -0.094\sigma ^{\frac{3}{2}})\times 10^{-6}  \nonumber
\label{fit4}
\end{eqnarray}%
where $\theta \equiv \frac{\delta \rho }{\rho }/\Delta _{c_{v}}$, $\sigma
\equiv a/a_{v}$. $\Delta _{c_{v}}$ and $a_{v}$ are 'input' parameters
defined when $\alpha (z_{v})=\alpha _{0}$ . 

The evolution of $\alpha \equiv \alpha _{b}$ in the background and inside
clusters depends mainly on the dominant equation of state of the universe
and the 
sign of the coupling constant $\frac{\zeta }{\omega },$ which is determined
by the theory and the dark matter identity. Here, we have assumed that $%
\omega =1$, so all the dependence is in $\zeta $. 
As shown in refs. \cite%
{bsm1}, $\alpha _{b}$ will be nearly constant for accelerated expansion and
also during radiation-domination far from the initial singularity (where the
kinetic term, $\rho _{\psi }$, can dominate). Evolution of $\alpha $ will
occur during the dust-dominated epoch, where $\alpha $ increases
logarithmically in time for $\zeta <0$. When $\zeta $ is negative, $\alpha $
will be a growing function of time but will fall for $\zeta $ positive \cite%
{bsm1}. 
Inside clusters $\alpha _{c}$ will have the same dependence on $\zeta $ as
it has in the background. 
The sign of $\zeta $ is determined by the physical character of the dark
matter that carries electromagnetic coupling: if it is dominated by magnetic
energy then $\zeta <0$, if not then $\zeta >0.$ Our numerical simulations
were performed assuming $\zeta =-2\times 10^{-4}$. If we had chosen $\zeta $
to be positive we would find that $\alpha _{b}$ would decrease as steeply as 
$z\rightarrow 0$. We choose the sign of $\zeta $ to be negative so $\alpha $
is a slowly growing function in time during dust domination. This is done in
order to match the latest observations which suggest that $\alpha $ had a
smaller value in the past \cite{murphy}. However, we know that on small
enough scales the dark matter will become dominated by a baryonic
contribution for which $\zeta >0.$ This will create distinctive behaviour
and will be investigated separately. Generally, past studies of spatially
homogeneous cosmologies have effectively matched the value of $\alpha _{b}$
with the terrestrial value of $\alpha $ measured today. However, it is clear
that the value of the fine structure constant on Earth, and most probably in
our local cluster, will differ from that in the background universe. This
feature has been ignored when comparing observations of $\alpha $ variations
from quasar absorption spectra \cite{murphy} with local bounds derived by
direct measurement \cite{prestage} or from Oklo \cite{oklo} and long-lived $%
\beta $-decayers \cite{beta}. A similar unwarranted assumption is generally
made when using solar system tests of general relativity to bound possible
time variations in $G$ in Brans-Dicke theory \cite{varyg}: there is no reason
why $\dot{G}/G$ should be the same in the solar system and on cosmological
scales. Since any varying-constant's model require the existence of
a scalar field coupled to the matter fields, 
our considerations apply to all other theories 
besides BSBM and to variations of other 'constants' \cite{other}. Note that
this feature is much less important when using early universe constraints
like the CMB or BBN \cite{cmbbbn}, since small perturbations in $\alpha$
will decay or become constant in the radiation era \cite{mota2}. 

In summary, using the BSBM theory we have shown that spatial variations in $%
\alpha $ will inevitably occur because of the development of large density
inhomogeneities in the universe. This was first shown in the linear regime,
when perturbations are small \cite{mota2}, and then $\frac{\delta \alpha }{%
\alpha }$ tracks $\frac{\delta \rho _{m}}{\rho _{m}}$ during the
dust-dominated era on scales smaller than the Hubble radius. We have used
the spherical collapse model to study the space-time variations in $\alpha $
in the non-linear regime. Strong differences arise between the value of $%
\alpha $ inside a cluster and its value in the background and also between
clusters. Variations in $\alpha $ depend on the matter density contrast of
the cluster region and the redshift at which it virialised. If the overdense
regions are still contracting and have not yet virialised, then the value of 
$\alpha $ within them will continue to change. Variations in $\alpha $ will
cease when the cluster virialises so long as it does so at moderate
redshift. This leads to larger values of $\alpha $ in the overdense regions
than in the cosmological background and means that time variations in $
\alpha $ will turn off in virialised overdensities even though they continue
in the background universe. 
The fact that local $\alpha $ values 'freeze in'
at virialisation, means we would observe no time or spatial variations in $%
\alpha $ on Earth, or elsewhere in our Galaxy, even though time-variations
in $\alpha $ might still be occurring on extragalactic scales. For a
cluster, the value of $\alpha $ today will be the value of $\alpha $ at the
virialisation time of the cluster. We should observe significant differences
in $\alpha $ only when comparing clusters which virialised at quite
different redshifts. Differences will arise within the same bound system
only if it has not reached viral equilibrium. Hence, variations in $\alpha $
using geochemical methods could easily give a value that is $100$ times
smaller than is inferred from quasar spectra.
We conclude that the
consideration of the evolution of inhomogeneities, notably the one inside
which we live, is essential if we are to make meaningful comparisons of
different pieces of astronomical and terrestrial evidence for the constancy
of $\alpha$.

\end{document}